\documentclass[reprint,12pt,NumberedRefs]{JASAnew}

\begin{document}

\title[]{A nonreciprocal and tunable active acoustic scatterer}

\author{Anis Maddi}
\email{anis.maddi@univ-lemans.fr}
\affiliation{Laboratoire d’Acoustique de l’Université du Mans (LAUM), UMR 6613,  Institut d’Acoustique - Graduate School (IA-GS), CNRS, Le Mans Université, France.}

 \author{Gaelle Poignand}
\affiliation{Laboratoire d’Acoustique de l’Université du Mans (LAUM), UMR 6613,  Institut d’Acoustique - Graduate School (IA-GS), CNRS, Le Mans Université, France.}

 \author{Vassos Achilleos}
\affiliation{Laboratoire d’Acoustique de l’Université du Mans (LAUM), UMR 6613,  Institut d’Acoustique - Graduate School (IA-GS), CNRS, Le Mans Université, France.}

 \author{Vincent Pagneux}
\affiliation{Laboratoire d’Acoustique de l’Université du Mans (LAUM), UMR 6613,  Institut d’Acoustique - Graduate School (IA-GS), CNRS, Le Mans Université, France.}

\author{Guillaume Penelet}			
\affiliation{Laboratoire d’Acoustique de l’Université du Mans (LAUM), UMR 6613,  Institut d’Acoustique - Graduate School (IA-GS), CNRS, Le Mans Université, France.}

\date{\today} 

\begin{abstract}
 
This paper is part of a special issue on Active and Tunable Acoustic Metamaterials.

A passive loudspeaker mounted in a duct acts as a reciprocal scatterer for plane waves impinging on either of its sides. However, the reciprocity can be broken by means of an asymmetric electroacoustic feedback which supplies to the loudspeaker a signal picked-up from a microphone facing only one of its sides. This simple modification offers new opportunities for the control and manipulation of sound waves. In this paper, we investigate the scattering features of  a pair of such actively controlled loudspeakers connected by means of a short and narrow duct. The theoretical and experimental results demonstrate that by tuning the feedback loops, the system exhibits several exotic effects, which include an asymmetric reflectionless configuration with one-way transmission or absorption, a directional amplifier with an isolation of $42$dB, and a quasi CPA-lasing configuration. All of these effects were achieved using a single setup in the subwavelength regime, highlighting the versatility of such an asymmetrically active scatterer.

\end{abstract}


\maketitle

\section{\label{sec:1} Introduction}


The control of sound waves is an active field of research, particularly driven by recent developments on acoustic metamaterials, which has widely extended the capacity to manipulate acoustic waves. Such materials can have various practical applications  \cite{cummer2016controlling,haberman2016acoustic}, including the absorption of sound waves \cite{ma2016acoustic,jiang2014ultra,yang2023acoustic,zhou2022broadband,huang2023sound,wang2024broadband}. To date, most developed systems retain reciprocity. However, a growing interest has emerged lately for the development of nonreciprocal acoustic materials. In contrast to conventional systems, these materials allow the unidirectional control of sound waves characterized by an asymmetric wave transmission.

In acoustics, the principle of reciprocity \cite{nassar2020nonreciprocity,rasmussen2021acoustic} can be broken by various methods, for instance by using large-amplitude waves to trigger nonlinear effects\cite{popa2014non,mojahed2019tunable,fu2018high,devaux2019acoustic,devaux2015asymmetric}, by imposing a temperature difference across a porous material \cite{maddi2023frozen,olivier2021nonreciprocal,biwa2016experimental,olivier2022asymmetric},  by modulating the medium properties in both space and time, or by using active control techniques \cite{shen2019nonreciprocal,zhu2020non,li2019nonreciprocal,chen2019nonreciprocal,geib2021tunable,penelet2021broadband,sasmal2020broadband,wen2023acoustic,guo2023observation,zhai2019active,padlewski2024amplitude}. Previous applications of these methods have led to the design of systems featuring transmission asymmetry, such as acoustic isolators and circulators\cite{fleury2014sound,pedergnana2024loss,wen2023acoustic,zhang2024nonreciprocal,mallejac2024experimental}. However, further studies are needed to cover all the potential applications, while also improving the design of non-reciprocal systems, especially considering their complexity for experimental implementations. Among the methods mentioned, active control allows for an easier manipulation of the acoustic waves \cite{fleury2015invisible,chen2018review,lissek2018toward,de2022effect,lissek2011electroacoustic,wang2024broadband,koutserimpas2019active,sergeev2023ultrabroadband,padlewski2023active}, but they can also lead to undesired instabilities (audio feedback).

In this paper, we introduce a tunable, subwavelength and nonreciprocal acoustic system which makes use of actively controlled loudspeakers. Non-reciprocity of a single scatterer is achieved by using a feedback loop, which consists of a microphone placed close to one side of a loudspeaker, and which feeds the latter with an amplified signal proportional to the pressure measured. The experimental system described in the following consists of two identical scatterers, each made up of a loudspeaker enclosed in a cavity and controlled with a feedback loop, while the two cavities are connected via a narrower duct. A single scatterer can be considered a building block of a non-Hermitian topological system, in which the emergence of the non-Hermitian skin effect and topological properties have been previously demonstrated \cite{maddi2024exact}. Nonhermitian topological systems\cite{okuma2023non}, as discussed by Ghaemi\cite{ghaemi2021compatibility}, can exhibit compelling scattering properties by controlling the nonreciprocal coupling in a periodic network, including coherent perfect absorption, lasing and reflectionless propagation. In this study, we demonstrate experimentally the possibility to build a versatile nonreciprocal acoustic scatterer with given geometry and components, by only tuning the gain in each independent feedback loop.

\begin{figure}
    \centering
    \includegraphics[width=0.48\textwidth]{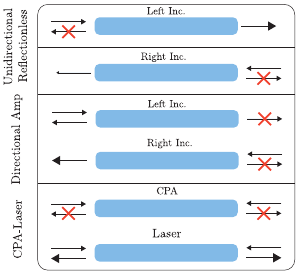}
    \caption{Overview of the effects explored in the present work. }
    \label{Summary}
\end{figure}

A theoretical description of a simplified version of the system is presented in section \ref{sec:2}, with an objective which is twofold. The first objective is to show that for one cell composed of a single actively controlled loudspeaker, it is possible to achieve a broadband non reciprocity by providing a gain in the feedback loop. It is also shown that by tuning two parameters, namely the  feedback gain and a change in cross-sectional area, the cell can act as a nonreciprocal asymmetric reflectionless two-port and an isolator. The second objective is to show that by using two coupled cells, the resulting two-port can keep the same nonreciprocal properties as for a unique cell,  while the tuning becomes more easy as it only requires adjusting the gain of each scatterer separately rather than a gain and the geometry. It is notably shown that the system can work as a directional amplifier\cite{malz2018quantum}, meaning that it not only acts as an isolator but also provides a transmission gain. Moreover, an additional effect of CPA-Laser \cite{auregan2017pt,poignand2021parity,longhi2010pt,chong2011pt,yang2024experimental} is enlightened, where the system can achieve either a strong amplification or a coherent perfect absorption of the input power at the same frequency, depending on the amplitude and phasing of the input waves. \color{black} A global picture of the effects explored in this work is presented in Fig.\ref{Summary}, and an experimental investigation of such exotic scattering properties is presented in Sec.\ref{sec:3}. 

\section{\label{sec:2} Theoretical description}

In this section, we first derive the governing equation of an active loudspeaker where we show the possibility of breaking acoustic reciprocity by supplying to the speaker a current proportional to the acoustic pressure on one side of the membrane. Then, we investigate an electroacoustic cell comprising a nonreciprocal active speaker placed inside a cavity, and we show that the introduction of a cross-sectional area change in addition to the adequate tuning of the gain allow to  control the scattering coefficients.  Finally, we explore the scattering features of a system comprising two electroacoustic cells, where the numerical results shows the possibility to suppress the reflection in a nonreciprocal system as well as to build as directional amplifier and a CPA-laser.

\subsection{breakdown of reciprocity}
\begin{figure}[ht]
    \centering
    \includegraphics[width=0.48\textwidth]{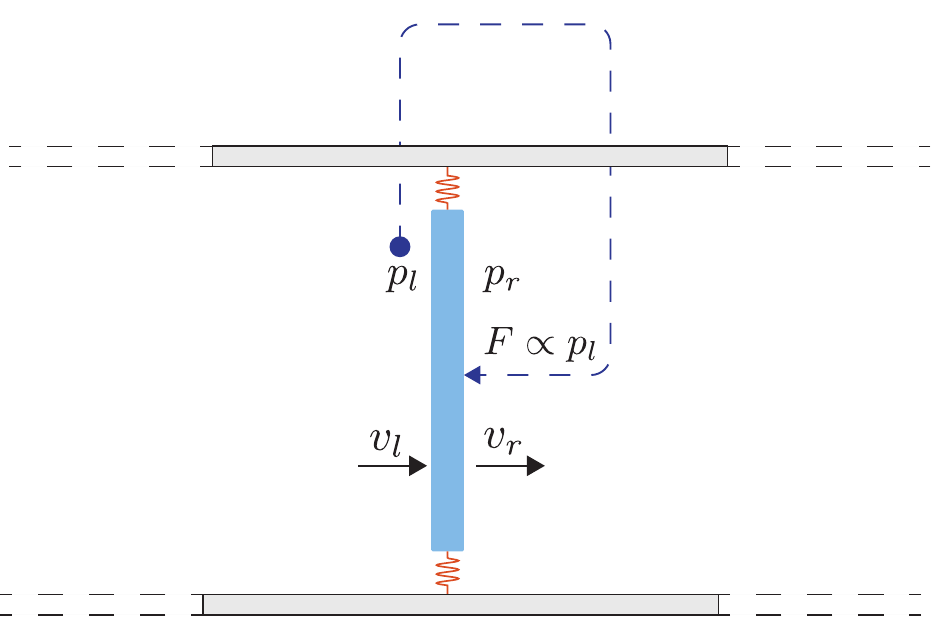}
    \caption{Schematic of the nonreciprocal electroacoustic device.}
    \label{cell1}
\end{figure}

A loudspeaker mounted inside a duct can be described as a mass-spring-damper system that is submitted to two external forces, namely a one caused by a pressure difference $p_l-p_r$ between the left and right sides of the speaker's membrane, and another one due to an (active) electrodynamic force $F$, as illustrated in Fig.\ref{cell1}. For a traditional moving coil loudspeaker, this force $F$ stems from the current $i$ that passes a coil of length $\ell$ in the presence of a magnetic field $B$. When an acoustic wave propagating along the duct axis arrives on a passive speaker (i.e., $F=0$), the latter just acts as a simple oscillator described as a rigid membrane with a mass $M_m$ and a surface area $S_m$, and a stiffness $K_m$ and  mechanical resistance $R_m$. The transmission/reflection/absorption of incident waves by the passive loudspeaker can be easily derived in terms of its scattering matrix. Now, if the speaker is active $F=B\ell i$, the electrodynamical force can be used to alter the acoustic field. The velocity $v$ of the loudspeaker's membrane is obtained by applying Newton's second law and writes as \cite{penelet2021broadband,guo2020improving},

\begin{align}
    Z_l v = (p_l-p_r) + \frac{B\ell}{S_m} i,   
    \label{Speaker_eq}
\end{align}
 \noindent
where $Z_l=\frac{1}{S_m}(R_m + j\omega M_m + \frac{K_m}{j\omega})$ is the impedance of the loudspeaker, $j^2=-1$, and  $B\ell$ is the electrodynamic force factor.

One way to control the speaker is to use an electroacoustic feedback loop, such that the loudspeaker is supplied with an electric current $i$ proportional to a pressure measured at one of its sides, creating an apparent symmetry. Herein, the microphone measures a pressure $p_l$, which corresponds to the pressure on the left-hand side of the loudspeaker (See Fig.\ref{cell1}). The signal detected by the microphone then passes through a amplifier with an adjustable gain $G$, such that the amplifier powers the loudspeaker with the current $i=Gp_l$.

\color{black}

Hence, Eq.\eqref{Speaker_eq} can be reformulated only in terms of the acoustic variables $(p,v)$. Moreover, by taking into account the continuity of the velocity $v$ between the left and right side of the membrane, i.e. $v=v_\ell=v_r$,  a transfer matrix $\mathbf{M_{0}}$ can be derived as follows:

\begin{align} 
\begin{pmatrix}  p_r\\ 
v_r  \end{pmatrix}=\underbrace{\begin{pmatrix}  t & -Z_{l}\\ 
0 & 1 \end{pmatrix}}_{\mathbf{M_{0}}}\begin{pmatrix}  p_l\\ 
v_l  \end{pmatrix}, 
\label{Map}
\end{align}
with
\begin{equation}
t=1-\frac{G B\ell}{S_m}.
\label{eq:def_t}
\end{equation}

It is worth noting that if the amplifier is active, $G\ne 0$, then the system becomes nonreciprocal, $\det(\mathbf{M_{0}})\ne 1$. Moreover, as the feedback loop includes only a static gain, the parameter $t$ is a real-valued and frequency independent coefficient. On the other hand, the loudspeaker operates as a passive resonator in the absence of a feedback loop ($G=0$).

\subsection{Scattering of one cell}

\begin{figure}[ht]
    \centering
    \includegraphics[width=0.48\textwidth]{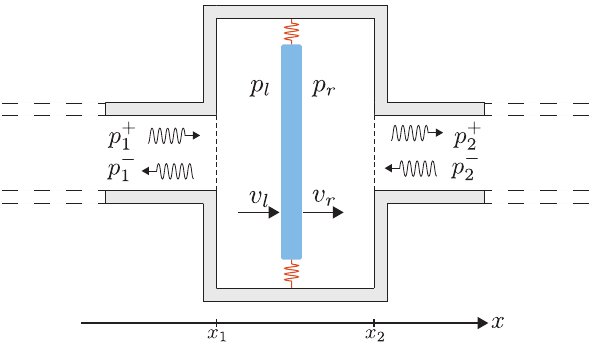}
    \caption{Schematic of the active electroacoustic element placed inside a cavity (of cross-sectional area $S_m$) that is connected to two narrower ducts (of cross-sectional area $S_d$).}
    \label{fig:enter-labegggl}
\end{figure}

In the following, a single cell is described, which is based on the feedback control mentioned above, but also includes a change in cross-sectional area on both sides of the loudspeaker, as described in Fig.\ref{fig:enter-labegggl}. 
The  system considered  consists of a speaker placed in a compact cavity of section $S_m$ that is connected to two narrower ducts of cross section $S_d$. As the system is considered to be compact, the continuity of both pressure and volume velocity applies at the interface between the duct and the cavity, such that 
\begin{subequations}
\begin{eqnarray}
    S_d v_1= S_m v_\ell, & S_d v_2 =S_m v_r  ,
    \\
    p_1 = p_\ell, & p_2=p_r.
\end{eqnarray}
\end{subequations}
By assuming that the pressure and the velocity are uniform in both cavities (owing to their compacity). The latter assumption will need to be revisited later for a more accurate description of the system.  

As a result, by taking into account the continuity of velocity and the pressure jump [see Eq.\eqref{Speaker_eq}] at the loudspeaker interface, the following equation can be obtained,
\begin{equation}
     (j\omega M_m +R_m +\frac{K_m}{j\omega})\frac{S_d}{S_m^2}v_1 =  (tp_1 - p_2). 
     \label{Eq.Imp}
\end{equation}

The pressures $p_{1,2}$ and velocities $v_{1,2}$  on both sides of the two-port can be decomposed in terms of forward $p^+$ and backward $p^-$ traveling waves as 
\begin{align}
    p= p^+ + p^- \,\, \, \text{\&}\,\, \, \rho cv= p^+ - p^-;
    \label{decomposition}
\end{align}
\noindent
where $c$ and $\rho$ are speed of sound and mean density of the fluid, respectively.

This decomposition is used to find the scattering matrix $\mathbf{S}$, which relates the ingoing waves $(p_1^+, p_2^- )$ to the outgoing waves $(p_1^-,p_2^+)$ in terms of the scattering coefficients. The scattering problem writes as
\color{black}

\begin{align} 
\begin{pmatrix}  p_2^+\\ 
p_1^-  \end{pmatrix}=\begin{pmatrix}  \mathbf{T}^+ &\mathbf{R}^-\\ 
\mathbf{R}^+ & \mathbf{T}^- \end{pmatrix}\begin{pmatrix}  p_1^+\\ 
p_2^-  \end{pmatrix},
\label{Smatrix}
\end{align}
 where the elements of the scattering matrix $\mathbf{T}^+$ and $\mathbf{R}^+$ ($ \mathbf{T}^-$ \& $\mathbf{R}^-$) represent the transmission and reflection coefficient of left (right) impinging waves,  respectively. Using the previous equations, these coefficients are expressed as follow,
 
 \begin{subequations}
\begin{align}
    \mathbf{T}^+&=\frac{\frac{2t}{\alpha}\frac{j\omega}{\omega_0}}{1-\frac{\omega^2}{\omega_0^2}+\left[\frac{1}{Q}+\frac{t+1}{\alpha}\right] \frac{j\omega}{\omega_0}},\label{Eq/T1} \\
    \mathbf{T}^-&=\frac{\frac{2}{\alpha}\frac{j\omega}{\omega_0}}{1-\frac{\omega^2}{\omega_0^2}+\left[\frac{1}{Q} +\frac{t+1}{\alpha}\right] \frac{j\omega}{\omega_0} }, \\
    \mathbf{R}^+&=\frac{1-\frac{\omega^2}{\omega_0^2}+\left[\frac{1}{Q}-\frac{t-1}{\alpha}    \right]\frac{j\omega}{\omega_0} }{1-\frac{\omega^2}{\omega_0^2}+\left[\frac{1}{Q}+\frac{t+1}{\alpha}\right]\frac{j\omega}{\omega_0} }, 
    \label{Eq.13_1}\\
     \mathbf{R}^-&=\frac{1-\frac{\omega^2}{\omega_0^2}+\left[\frac{1}{Q}+\frac{t-1}{\alpha}     \right]\frac{j\omega}{\omega_0} }{1-\frac{\omega^2}{\omega_0^2}+\left[\frac{1}{Q}+\frac{t+1}{\alpha}\right]\frac{j\omega}{\omega_0} }. 
     \label{Eq.14_1}
\end{align}
\label{Eq7}
\end{subequations}
where $\omega_0=\sqrt{\frac{K_m}{M_m}}$, $Q=\frac{K_m}{\omega_0 R_m}$ are the natural angular frequency and the quality factor of the mechanical resonator, and where $\alpha=\frac{K_m}{\omega_0\rho c S_m}\frac{S_d}{S_m}$ is a geometrical coupling parameter which accounts for the ratio of cross-sectional areas $S_m/S_d$.

Unsurprisingly, the expressions of the scattering coefficients show that as far as $t\ne 1$, the reciprocity is broken and $\mathbf{T^+}\ne \mathbf{T^-}$. Moreover, as all the coefficients depend on the gain $t$ and on the change in cross-section through the coupling parameter $\alpha$, this opens the way for adjusting these two parameters to obtain interesting scattering effects.  

Following Figure \ref{Summary}, a first objective can be, for instance, to make the two-port reflectionless from one side, which can be achieved by setting the numerator to zero in Eq.\eqref{Eq.13_1} such that $\mathbf{R}^{+}$=0, or in \ref{Eq.14_1} such that $\mathbf{R}^{-}$=0. Hence, by setting $\mathbf{R^{\pm}}=0$, we obtain the following complex valued equation,
\begin{equation}
    {\left[\frac{1}{Q}\pm\frac{t-1}{\alpha}     \right]\frac{j\omega}{\omega_0}} +{1-\frac{\omega^2}{\omega_0^2}} =0.
    \label{complex-Eq}
\end{equation}
The solutions of Eq.\eqref{complex-Eq} are found by solving the real and imaginary parts separately, leading to
\begin{subequations}
\begin{align}
    t&=1\pm\frac{\alpha}{Q}, \label{NW1} \\
    \omega &=\omega_0. \label{NW2}
\end{align}
    \label{LOLL}
\end{subequations}
As a result, the two-port can be made one-sided reflectionless at the angular frequency $\omega_0$ by adjusting either the gain $t$ or the coupling parameter $\alpha$ to satisfy Eq.\eqref{NW1}.

Another objective can be to make the system transmissionless from one side (See fig.\ref{Summary}), and Eq.\eqref{Eq/T1} shows the possibility to block the transmission of left incident waves by setting $t=0$ such that $\mathbf{T^+}=0$. Interestingly, the latter broadband suppression of the transmitted waves is achieved together with a broadband unitary reflection $\mathbf{R^+}=1$, as can be shown by setting $t=0$ in Eq. \eqref{Eq7}. These broadband effects are obtained no matter the choice of the coupling parameter $\alpha$, such that adjusting this second parameter can help controlling reflection and transmission from the other side. In particular,   $\alpha$ can be chosen such that scatterer becomes reflectionless for right incident waves at the angular frequency $\omega_0$,  leading to the following scattering matrix,

\begin{align} 
\mathbf{S}(\omega_0)=\begin{pmatrix}  0 &0 \\ 
1 & 1 \end{pmatrix},
\label{SLmatrix_didoe}
\end{align}
with $t=0$ and $\alpha=Q$.

This scattering matrix describes an acoustic isolator, in which the two-port transmits only right incident waves, whereas it is fully reflective from the opposite side.

The two examples above show that such a system can provide some interesting nonreciprocal scattering effects, obtained by adjusting the gain $t$ and the parameter $\alpha$. However, there exists limitations for the practical implementation of this system. A first limitation is the fact that, contrarily to the gain $t$, the parameter $\alpha$ is a geometrical parameter which depends on the ratio $S_d/S_m$, and it cannot be tuned for a single device. Moreover, the model presented above is a simplified model, which notably ignores the presence of a cavity on both sides of the membrane. Actually those cavities mostly act as additional compliances and inertances which impact the design rules for achieving both reflectionless or transmissionless configurations, as highlighted in the Appendix where a more accurate model is presented.

In order to have a versatile acoustic scatterer with given geometry and components, we decided to introduce a second electroacoustic cell which is connected to the first one. This additional cell serves as an extra degree of freedom to substitute the parameter $\alpha$ which is typically fixed. Therefore, if one fixes a geometry and a loudspeaker (i.e. $\alpha, Q, \omega_0$ are set), the new parameters to tune are the independent gains $t_1$ and $t_2$ of the two electroacoustic cells. 

\subsection{Scattering of two cells}
\begin{figure}[ht]
    \centering
    \includegraphics[width=0.48\textwidth]{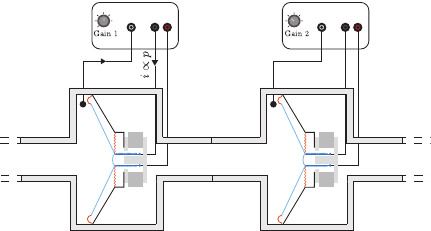}
    \caption{Sketch of the system comprising two electroacoustic cells. This setup corresponds to the one used experimentally in this study. \color{black}}
    \label{Exp}
\end{figure}

\paragraph{Simplified model}
As mentioned previously, while a single cell can be used (in principle) to break the reciprocity and to control either the reflection or the transmission by tuning the gain $t$ and choosing a parameter $\alpha$, it is preferable to consider a system consisting of two cells, so that the additional gain offers a tunable degree of freedom. In the following, we first consider a simplified description of a system consisting of two unit cells separated by a duct. This two cell configuration is the one that will be used in experiments, as shown in Fig.\ref{Exp} which gives a sketch of the experimental setup. The two active loudspeakers can be controlled independently, thus offering greater flexibility to adjust the scattering coefficients.

Similarly to the case of a single cell, we can derive simplified expressions of the scattering coefficients by omitting the impact of the cavities. Here, if one adjusts the length $L_d$ of the connecting duct such that it corresponds to $\pi c/\omega_0$, namely one half of the wavelength at the resonance frequency of the loudspeakers, then the scattering matrix at $\omega=\omega_0$ is given by,

\begin{align} 
\mathbf{S}(\omega_0)=\frac{1}{\frac{\alpha}{Q}+\frac{t_1t_2+1}{t_2+1}}\begin{pmatrix}  2t_1t_2 & \frac{\alpha}{Q}+\frac{t_1t_2-1}{t_2+1} \\ 
\frac{\alpha}{Q}-\frac{t_1t_2-1}{t_2+1} & 2 \end{pmatrix}.
\end{align}

All the scattering coefficients are explicitly dependent on the gains $t_1$ and $t_2$ of each speaker. This means that once a geometry and a loudspeaker are selected  (i.e., $ \omega_0$, $Q$ and $\alpha$ are fixed), the scattering properties of the system can still be adjusted with the two parameters $t_1$ and $t_2$, which can be easily tuned through the gains of the feedback loops.

  \begin{figure}[ht]
    \centering
    \includegraphics[width=0.48\textwidth]{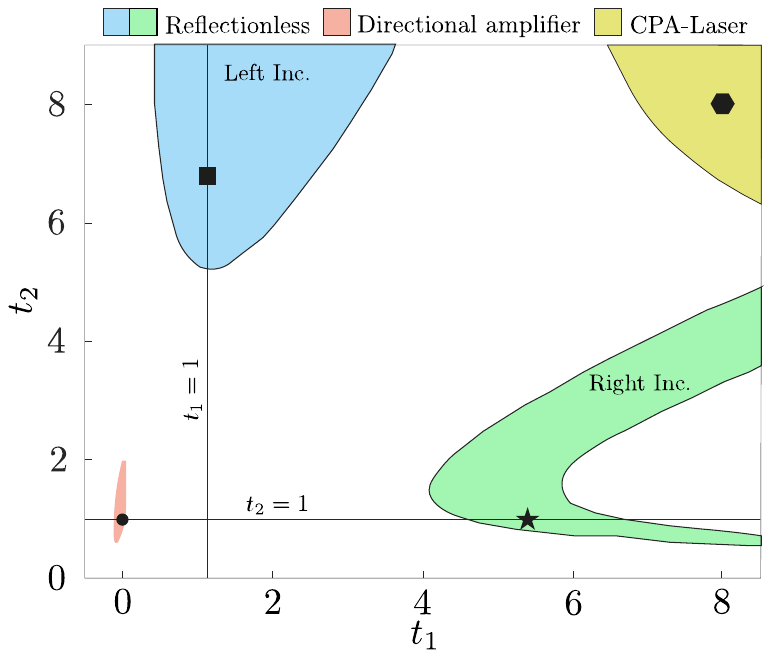}
    \caption{Map of the effects as function of the parameters $t_1$ and $t_2$. The markers (star, square, dot and polygon) represent the experimental configurations  presented in Sec.\ref{sec:3}. }
    \label{effect}
\end{figure}

This simplified expression of the scattering matrix coefficients indicates that some effects similar to the ones discussed for one cell can be obtained. This is achieved by adjusting $t_1$ and $t_2$, as opposed to $t$ and $\alpha$ in the single cell case. For instance, if $t_1=0$ and $1+t_2=Q/\alpha$ then $\mathbf{T}^+=0$, $\mathbf{R}^+=1$, $\mathbf{R}^-=0$ and $\mathbf{T}^-=Q/\alpha$. Yet, this analytical approach is limited since it does not account for the impact of the cavities. Therefore, a numerical approach was adopted for the analysis of the scattering, using the transfer matrix method presented in the Appendix.

\paragraph{Numerical results}
We now proceed to a numerical study of the scattering properties of the two-port as functions of the frequency and for different values of the gains $t_1$ and $t_2$. Calculations are performed from a more accurate model which is described in the Appendix, and accounts for the impact of the cavities which were omitted in the simplified model above. Herein, the aim is to explore the effects discussed previously with values taken from the experimental apparatus which is described in the next section. More precisely, three distinct effects are investigated and a mapping of their occurrence is given in Fig.\ref{effect}, depending on the choices of $t_{1,2}$ . The different configurations and their corresponding effects are categorized as follows :

\begin{itemize}
    \item[$\star$] 
     \textbf{One-sided Reflectionless.} In this configuration, at least one of the reflection coefficients approaches zero, $\mathbf{R^{\pm}}\xrightarrow[]{}0$. The numerical results are highlighted with blue or green surfaces on the map, and correspond either to a vanishing $\mathbf{R^{+}}$ or a vanishing $\mathbf{R^{-}}$, respectively. In practice, the system cannot be perfectly reflectionless, so the map is generated from the definition of a criterion. Herein, the system is referred to as reflectionless if the magnitude of a reflection coefficient is less than  $\mathbf{R^{\pm}}<0.05$.

\item[$\star$] \textbf{Directional amplifier.} This particular system act as an isolator coupled to an amplifier, in which the transmissive port has a gain $A>1$. In the ideal case, the scattering matrix looks as follows:
\begin{equation}
    \mathbf{S}=\begin{pmatrix}  0 &0 \\ 
1 & A \end{pmatrix}.
\end{equation}

The configurations that allow to obtain such a system are mapped with the pink surface in Fig.\ref{effect}. It was obtained numerically by imposing conditions on the scattering coefficients such that the right incident waves are transmitted with a gain $\mathbf{T^{-}}>1$ and minimal reflection $\mathbf{R^-}<0.1$. On the opposite propagation direction, the magnitude of the transmission need to be minimized while the reflection should be around unity. These conditions can be satisfied by imposing a restriction on the determinant of the transfer matrix of the system $\mathbf{M}$ such that $\vert \det(\mathbf{M}) \vert<0.1$. The resulting minimal isolation factor is $20\log(\mathbf{\vert \frac{T^-}{T^+}}\vert)>20$ dB.

\item[$\star$]  \textbf{CPA-Laser.} Another effect is the CPA-lasing which can be investigated from the singular values $\sigma_{\pm}$ of the scattering matrix. The singular value decomposition satisfies,

    \begin{equation}
        \mathbf{S}=\mathbf{U} \mathbf{\Sigma} \mathbf{V^\dagger},
    \end{equation}
where $^\dagger$ stands for the conjugate transpose, and where $\mathbf{U}$ and $\mathbf{V}$ are the orthonormal left and right singular vectors, respectively. Two vectors $\mathbf{U}_{\pm}$ and $\mathbf{V}_{\pm}$, are associated to each singular value $\sigma_{+}$ and $\sigma_{-}$, such that $\mathbf{\Sigma}=\diag(\sigma_+,\sigma_-)$. The singular value decomposition also satisfies
\begin{equation}
    \mathbf{S}\mathbf{V}_{\pm}=\sigma_{\pm} \mathbf{U}_{\pm}.
\end{equation}
Hence, the singular values decomposition (SVD) allows to linearly map the input waves $\mathbf{V_{\pm}}$  to the output waves $\mathbf{U}_{\pm}$ with a scaling factor $\sigma_{\pm}$ \cite{guo2023singular}. The SVD is useful for quantifying the dissipation or generation of acoustic power $\mathcal{P}$ \cite{auregan1999determination,holzinger2014optimizing}, as it allows to find the bounds of the power ratio  between the input and output, such that
\begin{eqnarray}
    \sigma_-^2\le\frac{\mathcal{P}_{out}} {{P}_{in}}=\frac{\vert p_2^+\vert^2 + \vert p_1^-\vert^2}{\vert p_1^+\vert^2 + \vert p_2^-\vert^2}\le  \sigma_+^2.
\end{eqnarray}

As a result, if the scatterer has a vanishing singular value $\sigma_- = 0$ and if the incidents waves are adjusted according to the right singular vector $\mathbf{V}_-$ (i.e., if the input vector  $[p_1^+,p_2^-]^T \propto \mathbf{V}_-$), then the system operates as a coherent perfect absorber (CPA), meaning that the incident power is fully absorbed by the system. Inversely, lasing happens when the singular value $\sigma_+$ is greater than unity, which means that the scatterer can amplify the incident acoustic power with a maximum amplification of $\sigma_+^2$ if the input waves are tuned according to the right singular vector $\mathbf{V}_+$. Finally, a CPA-Laser configuration is observed if both coherent perfection absorption and lasing can be achieved at the same frequency.

In Figure \ref{effect}, the configurations providing CPA-lasing are highlighted with the yellow color on the map. It was considered here that CPA-laser is observed if two conditions are satisfied, namely an upper bound for the CPA $\sigma_-<0.1$ and   a lower bound for the lasing $\sigma_+>5$. 

\end{itemize}


 Overall, the mapping provided in Fig.\ref{effect} gives information on the different settings possible by tuning the parameters $t_1$ and $t_2$. It notably shows that the one-sided reflectionless state tends to occur when the nonreciprocal parameters $t_{1,2}$ are distinct, since it enhances the mirror asymmetry.  Meanwhile, the results also confirm that directional amplifier configuration is located nearby the vertical line $t_1=0$ as it allows to fully block left incident waves. Finally, CPA-laser configurations are likely to occur at higher transmission asymmetries $t_1t_2>>1$. 
 
 In the next section, an experimental demonstration of each effect is provided, with the selected experimental configurations illustrated in this map with the black markers.

\section{\label{sec:3} Experimental Results}

In this section, the experimental scattering coefficients of the system are measured as functions of the frequency, and compared with numerical results. The  scattering matrix coefficients are measured using an impedance sensor method (see \cite{macaluso2011trumpet,bannwart2013measurements} for a description of the method). Note that the system remains stable over a wide range of values for $t_{1,2}$, which allowed us to explore several configurations. However, for large values of the gains, i.e. $t_{1,2}>8$, or for negative values $t_{1,2}<0$, undesired self-sustained oscillations are triggered in the system.

The experimental system consists of two unit cells as illustrated in Fig.\ref{Exp}, each with its electroacoustic feedback loop composed of a loudspeaker (model Aura NSW2) with a  resonance frequency estimated at $268$Hz, a microphone (Bruel \& Kjaer, model 4938), and a current amplifier. Each cell consists of a cavity of length $L_c=1.8 \text{cm}$ and cross-section $S_c=15 \text{cm}^2$, connected on both sides to a duct of length  $L_{d}=9 \text{cm}$ and a cross-section $S_{d}=0.5\text{cm}^2$. The total length of the system is $L_{tot}\approx 0.22$m. Most of the effects discussed below are observed at a frequency of about 200 Hz, corresponding to a wavelength of about 1.7 m. Hence, the typical length of the scatterer is much lower than the wavelength. 

\subsection{One-sided reflectionless}

\noindent\paragraph{One-way amplification}

\begin{figure}[ht]
    \centering
    \includegraphics[width=0.48\textwidth]{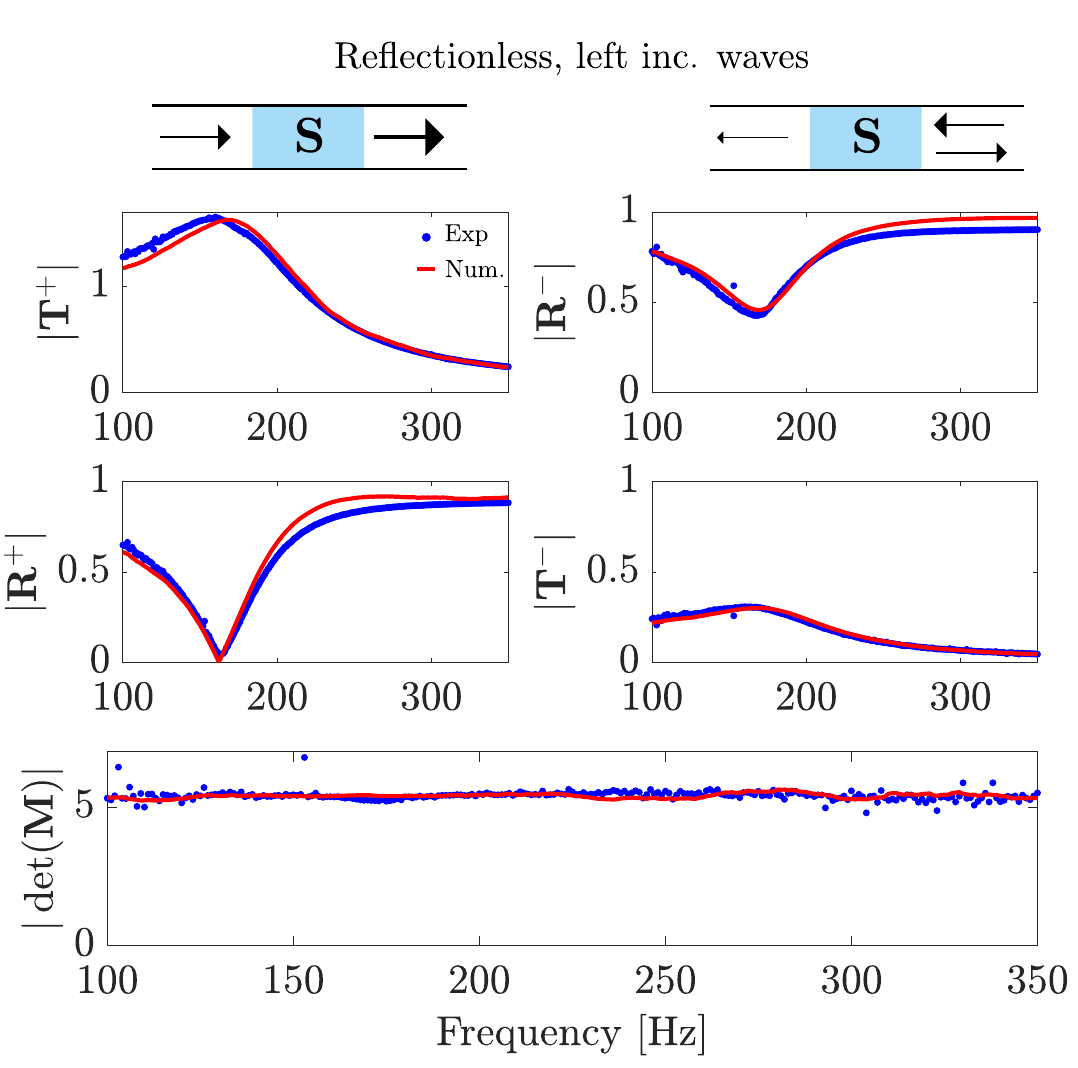}
    \caption{ Magnitude of the scattering coefficients and determinant of the transfer matrix as functions of the frequency for $t_1=1$ and $t_2=5.4$. Blue and red colors represent the experimental and theoretical results, respectively. }
    \label{Reflectionless_left}
\end{figure}

First, the one-sided  reflectionless configuration is investigated, which is highlighted in Fig.\ref{effect} by the star and square markers. The scattering coefficients and transfer matrix determinant are plotted in Fig.\ref{Reflectionless_left} as functions of the frequency for $t_1=1$ and $t_2=5.4$ (star marker). The blue and red lines represent the experimental and the theoretical results, respectively. In this configuration, the experimental reflection coefficient for a left-incident wave reaches $\vert \mathbf{R}^+\vert=0.04$ at a frequency of $f=163 \text{Hz}$, which corresponds to wavelength $\lambda=9L_{tot}$. Meanwhile, the transmission coefficient for a wave incident from the left reaches $\vert \mathbf{T}^+ \vert \approx 1.64$, indicating an unidirectional amplification of the left incident waves. In contrast, for a right impinging wave, the scatterer acts essentially as an absorber, with a reflection of $\vert \mathbf{R}^-\vert\approx 0.47$ and a transmission of $\vert \mathbf{T}^-\vert\approx 0.3$. Consequently, the absorption coefficient is $\alpha^-=1-\vert \mathbf{T^-} \vert^2 -\vert \mathbf{R^-}\vert =0.69$. Moreover, the determinant of the transfer matrix is almost constant over the whole frequency range, with $\det(\mathbf{M}) \approx 5.4$, indicating a constant asymmetry between the transmission coefficients.

\begin{figure}[ht]
    \centering
    \includegraphics[width=0.48\textwidth]{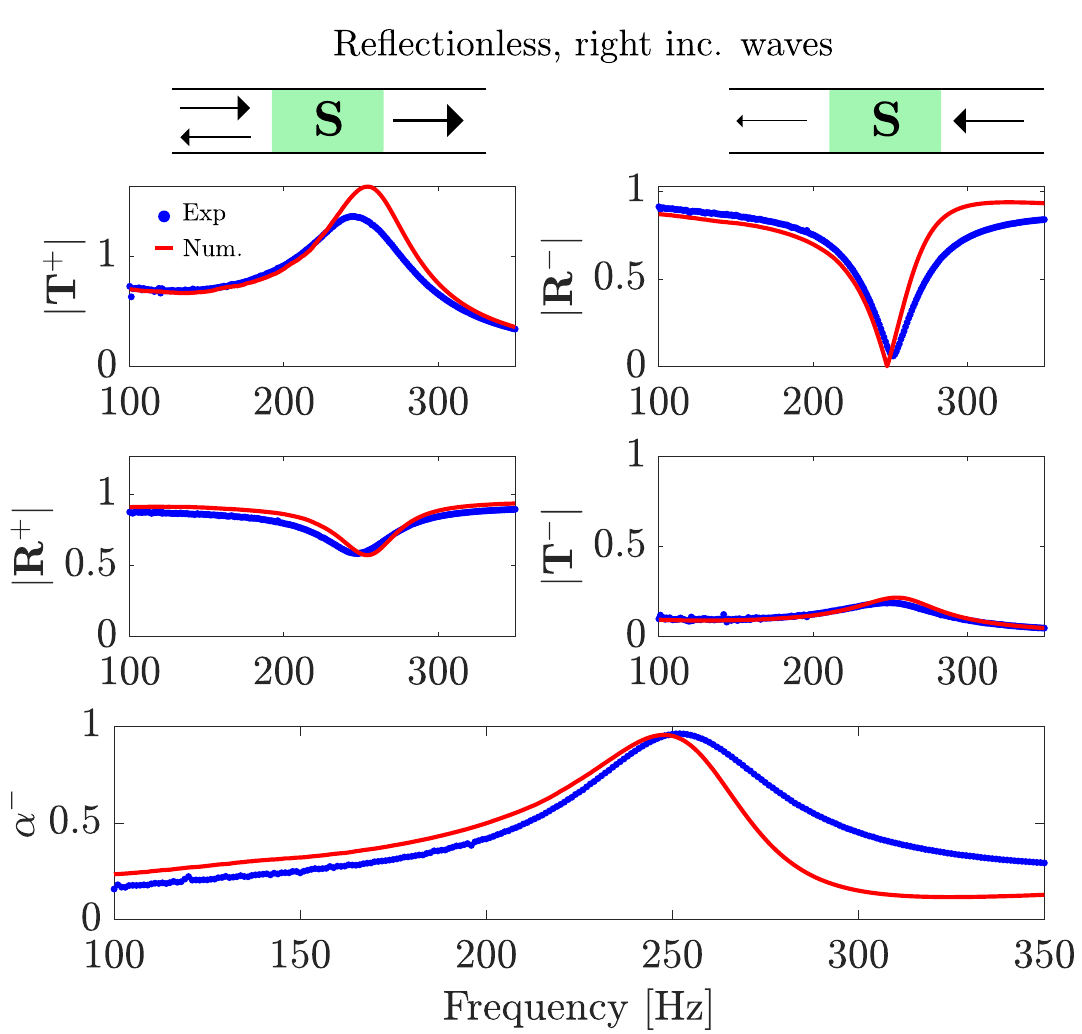}
    \caption{ Magnitude of the scattering coefficients and absorption coefficient $\alpha^-$ as a function of the frequency for $t_1=6.4$ and $t_2=1$. Blue and red colors represent respectively the experimental and theoretical results.}
    \label{Reflectionless_right}
\end{figure}

\noindent\paragraph{One-way absorption}

Having achieved a reflectionless configuration for left-hand impinging waves, the configuration yielding a reflectionless state for wave incident on the right side is next investigated. Figure \ref{Reflectionless_right} depicts the magnitudes of the scattering coefficients as functions of the frequency for $t_1=6.4$ and $t_2=1$. For this experimental setting, a wave incident from the right-hand side is mostly absorbed, resulting in a reflection coefficient of $\mathbf{R}^-\approx 0.06$ and a transmission coefficient of $\mathbf{T}^-\approx 0.21$ at $f=252 \,\text{Hz}$. Meanwhile, for a leftward incident wave, the system  has transmission higher than unity and a high reflection coefficient. In this configuration, the system achieves a high subwavelength absorption of the right incident waves with $\alpha^-\approx 0.96$ and $L_{tot} <\lambda/6$.

\subsection{Directional amplifier}

\begin{figure}[ht]
    \centering
    \includegraphics[width=0.48\textwidth]{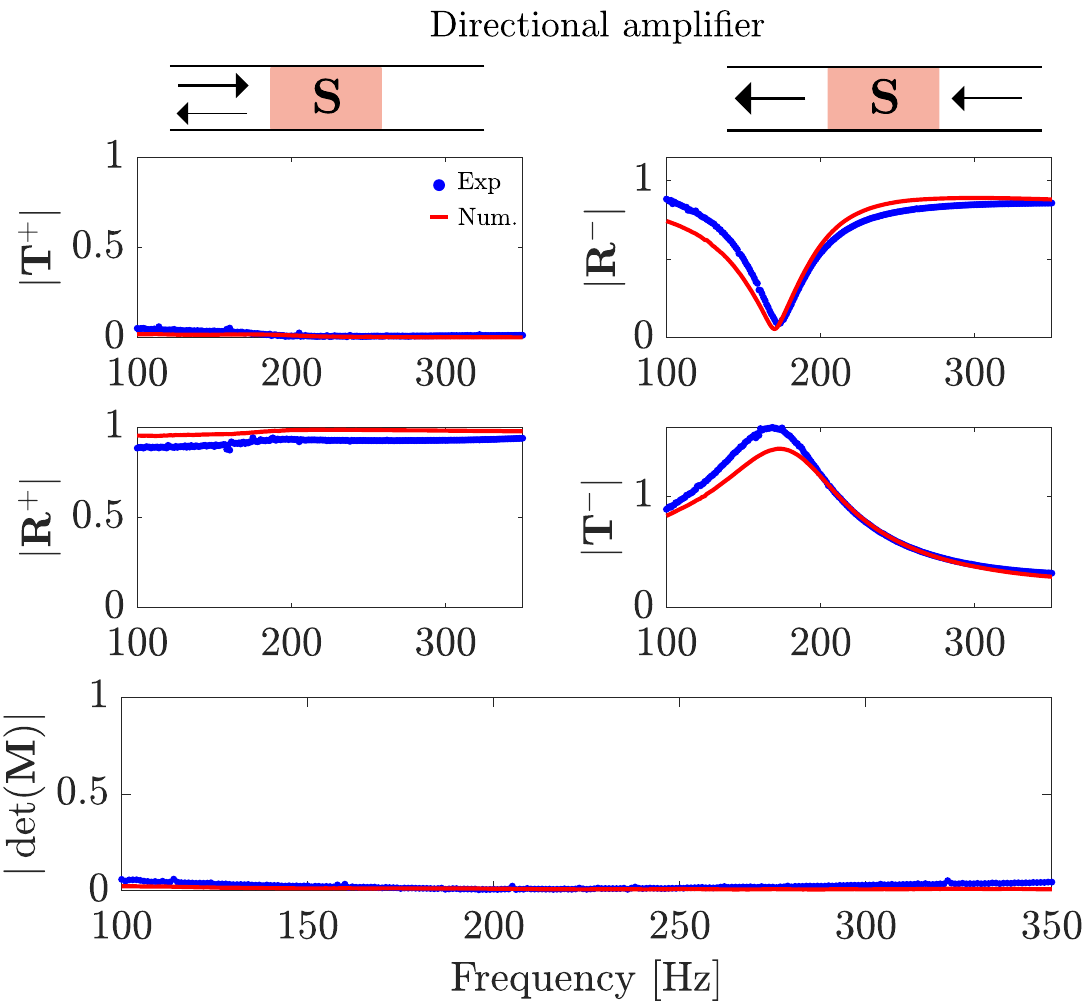}
    \caption{Magnitude of the Scattering coefficients and determinant of the transfer matrix as a function of the frequency for $t_1=0 $ and $t_2=1$. Blue and red colors represent respectively the experimental and theoretical results.}
    \label{diode}
\end{figure}

Next, a directional amplifier configuration is investigated experimentally, which is denoted in the map of Fig.\ref{effect} with the dot marker. In Fig.\ref{diode} the magnitude of the scattering coefficients and the determinant of the transfer matrix are shown as functions of the frequency for $t_1\approx 0$ and $t_2=1$. In this configuration, the loudspeaker gain $t_1$ is adjusted to achieve a broadband zero determinant, $\det(\mathbf{M})\approx 0$, which results in a negligible transmission coefficient for a leftward incident wave, $\mathbf{T}^+\approx 0$, as well as a high and broadband isolation factor of $42$dB. Moreover, the left-sided reflection is close to unity, indicating that the device acts as a rigid wall for left incident waves. Meanwhile, for a right-sided incident wave, the system allows a high transmission at low frequency, particularly around $f=170 \text{Hz}$ with an amplification of $A=1.62$. At this frequency, the reflection coefficient goes down to less than 10 \%. Hence, at $f=170$ Hz, the system acts as an amplifying diode by transmitting only waves incoming from the right side.

\subsection{Coherent-Perfect Absorption and Lasing}

\begin{figure}[ht]
   \centering
    \includegraphics[width=0.48\textwidth]{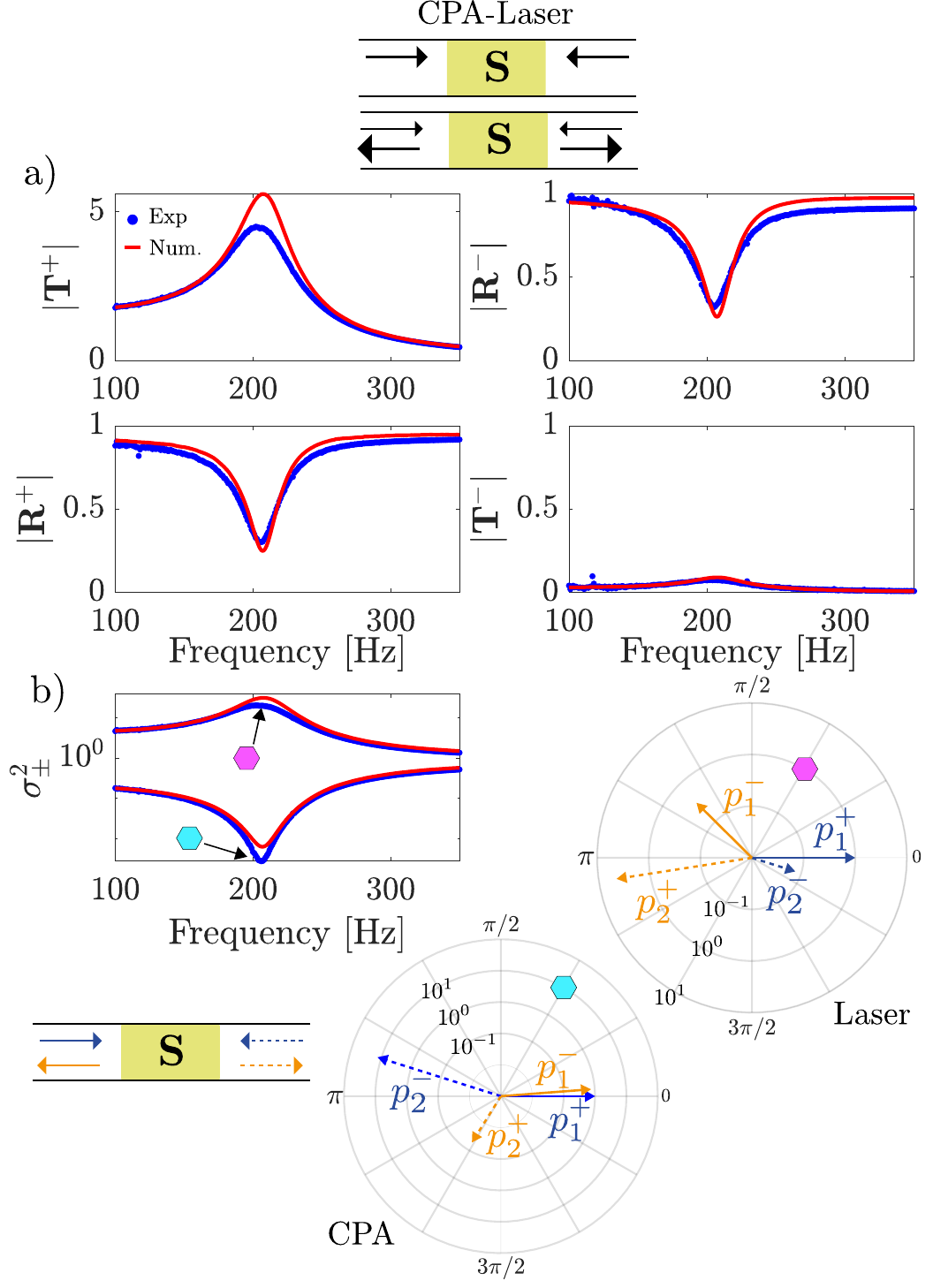}
    \caption{(a) Magnitude of the scattering coefficients as a function of the frequency for $t_1=t_2=8$. Blue and red colors represent the experimental and theoretical results, respectively. (b) Singular values $\sigma_{\pm}^2$ and the associated singular vectors at the CPA-Lasing configuration.}
    \label{CPA_laser}
\end{figure}

Finally the CPA-laser mode is investigated experimentally, with a configuration marked by the polygon in Fig.\ref{effect}. The magnitude of the  scattering coefficients and the singular values are plotted in Fig.\ref{CPA_laser} as functions of the frequency. Here, the gain of both cells was adjusted to the largest possible value before triggering an instability, corresponding to $t_1=t_2\approx 8$. In such a configuration, the system is characterized by a high and broadband isolation factor $20\log(\frac{\mathbf{T^+}}{\mathbf{T^-}})\approx 36$, with a strong amplification of left incident waves $\mathbf{T}^+ \approx 4.5$ and a low transmission from the opposite side. Furthermore, the system is mirror-symmetrical in this case, and the reflection coefficients reach a minimal value of $0.2$ at $f=205 \text{Hz}$.

The squares of the singular values $\sigma_\pm^2$, which quantify the scattered acoustic power, are also plotted as functions of the frequency. It is notably found that around $f=205$ Hz, the singular values reach two extreme values, such that the singular value associated to the coherent perfect absorption is $\sigma_-^2\approx 0.0025$ while the one for lasing $\sigma_+^2\approx 20.25$. These two states of large absorption or large amplification can be achieved by tuning the input waves accordingly with the singular vectors $\mathbf{V}_{\pm}$. Due to the strong nonreciprocity of the system, these eigenvectors correspond primarily to waves incident on only one side. Lasing occurs when the input waves satisfy $\vert p_1^+ \vert =13.2\vert p_2^- \vert$, denoting a predominance of left-incident waves, while CPA is observed when the input waves satisfy $\vert p_1^+ \vert =(13.2)^{-1}\vert p_2^- \vert$, thus corresponding mostly to a right-incident wave. As with the input, the output waves are predominantly one sided, such that the lasing is predominantly to the right $\vert p_2^+ \vert=13.9\vert p_1^- \vert$, meanwhile the remaining unabsorbed power for the CPA is to the left.

\section{Conclusion}

Using a simple nonreciprocal acoustic device, composed of two actively controlled loudspeakers with asymmetrical feedback loops, we have demonstrated the possibility to exhibit various scattering effects by simply tuning the feedback gain of each loudspeaker. The experimental and numerical results show good agreement as well as a large range of scattering effects, including a nonreciprocal reflectionless propagation with a one way transmission gain of $1.64$ or an absorption of $96\%$, a directional amplifier, and a nonreciprocal CPA-Laser configurations. The presented device yields a broadband and high isolation up to $42$ dB and operates in the subwavelength regime, typically with $ L_{tot} \sim \lambda/6$ or less for the effects highlighted here in experiments. The electroacoustic scatterer studied has the advantage of being simple to implement and to tune, as it only requires to adjust the gains of the amplifiers. In future works, it could also be interesting to use digital controllers, as they offer more advanced possibilities in terms of electroacoustic feedbackloops.



\section*{Acknowledgments}
\color{black}
V.A. acknowledges financial support from the NoHeNA project funded under the program Etoiles Montantes of the Region Pays de la Loire. V.A. Is supported by the EU H2020 ERC StG "NASA" Grant Agreement No. 101077954
\color{black}

\section*{Author Declarations}
\subsection*{Conflict of Interest}
The authors have no conflicts of interest to declare.

\section*{Data Availability}
The data that support the findings of this study are available on request from the corresponding author.

\appendix 

\section{Transfer matrix}
\subsection{Full model for one cell}
The transfer matrix of a straight duct of length $L_{c}$ is given by,
\begin{equation}
    \mathbf{M}_{c}=\begin{pmatrix}  \cos(kL_c)& -i\rho c \sin(kL_c)\\ 
-i\sin(kL_{c})/(\rho c) & \cos(kL_{c}) \end{pmatrix}.
\end{equation}

At the interface between two ducts with a different cross-section, the continuity of pressure and volume velocity allows to write the following transfer matrix 
\begin{equation}
    \mathbf{M}_{S_1-S_2}=\begin{pmatrix}  1&0\\ 
0 & \frac{S_1}{S_2} \end{pmatrix},
\end{equation}
for a transition from a cross section $S_1$ to $S_2$, A speaker with a cross-section $S_m$ and a transfer matrix $\mathbf{M_0}$ is enclosed in a cavity of cross-section $S_c$ connected on both sides to ducts with a cross-section $S_d$. The transfer matrix $\mathbf{M}_{\text{cell}}$ that describes this cell is obtained by using the continuity of pressure and velocity at each interface, which results in the following matrix

\begin{equation}
    \mathbf{M}_{\text{cell}} = \mathbf{M}_{S_c-S_d} \mathbf{M}_{\text{c}} \mathbf{M}_{S_m-S_c} \mathbf{M_0} \mathbf{M}_{S_c-S_m} \mathbf{M}_\text{c} \mathbf{M}_{S_d-S_c}. \label{one_cell}
\end{equation}

\subsection{Simplified version}\label{compliance}

Eq. \eqref{one_cell} can be simplified by assuming that the speaker and the cavity have the same cross-section, $S_m=S_c$. Additionally, when the cavity is short compared to the typical wavelength, we can further simplify the problem by taking into account the acoustic compliance $C=L_c S_c/(\rho c^2)$  and inductance $I=\rho L_c/S_c$ of the cavity, such that the transfer matrix writes as,

\begin{equation}
    \mathbf{M}_{c}=\begin{pmatrix}  1& -j\omega I\\ 
-j\omega C & 1 \end{pmatrix}.
\end{equation}
 
After  some algebra, one finds the following condition on the frequency to achieve a vanishing reflection coefficient $\mathbf{R^{\pm}}$,
 \begin{equation}
      \omega=\frac{\omega_1}{\sqrt{1-\omega_1^2\frac{C(t+1)}{\omega_0\alpha}\frac{\rho c S_m}{S_d}}},
      \label{eq:appen_reflectionless_cond}
 \end{equation}
\noindent
where $\omega_1=\sqrt{\frac{K}{M+IS_m}}$.  

Equation \eqref{eq:appen_reflectionless_cond} shows that the angular frequency at which the reflection can be suppressed depends on the compliance and inductance of the cavity, and that it might not have real solutions for all configurations.

\subsection{Transfer matrix of $2$ cell}\label{Appendix B}

Starting with equation \eqref{one_cell} that allows to write the transfer matrix of one cell, we can write the transfer matrix $\mathbf{M}_{sys}$ of a system composed of two cells connected by a duct of transfer matrix $\mathbf{M}_c$. The resulting transfer matrix is given by the following product,
\begin{equation}
    \mathbf{M}_{\text{sys}}= \mathbf{M}_{\text{cell},2} \mathbf{M}_{\text{c}}\mathbf{M}_{\text{cell},1},
\end{equation}
\noindent
where $\mathbf{M}_{\text{cell},1}$ and $\mathbf{M}_{\text{cell},2}$ stand for the transfer matrix of one cell with a gain $t_1$ or $t_2$.
\color{black}

\bibliography{sampbib}

\end{document}